\documentclass[doublecol]{epl2}
\usepackage{amssymb}

\input{tcilatex}
\title{Possible strong electron-lattice interaction and \\
giant magneto-elastic effects in Fe-pnictides}
\shorttitle{Possible strong electron-lattice interaction}
\author{\textsc{Miodrag L.\ Kuli\'{c}}\inst{1} \and \textsc{Amir A.
Haghighirad}\inst{2}} \shortauthor{M.L.\ Kuli\'{c} \etal}
\institute{\inst{1} Goethe-University Frankfurt, Theoretical
Physics,
60054 Frankfurt/Main, Germany \\
\inst{2} Goethe-University Frankfurt, Institute of Physics, 60054
Frankfurt/Main, Germany} 
 \abstract {The possibility for an appreciable
many-body contribution to the electron-phonon interaction ($EPI$)
in Fe-pnictides is discussed in the model where $EPI$ is due to
the electronic polarization of $As$- ions. The polarization
induced $EPI_{pol}$ potential $V_{ep}$ is large for vibrations of
the $As$-ions and depends strongly on the As-Fe distance $d$, i.e.
$V_{ep}\sim d^{-4}$. The $EPI_{pol}$ coupling $g_{ep}^{pol}$ is
much larger than the one obtained in the $LDA$ band structure
calculations, with $g_{ep}^{pol}(\sim 16$ $eV/\mathring{A})\gg
g_{ep}^{(LDA)}(<1$ $eV/\mathring{A})$ and the bare pairing
$EPI_{pol}$ coupling constant $\lambda _{ep,A_{1g}}^{0}\sim 1$. It
contributes significantly to the intra-band s-wave pairing and an
appreciable positive As-isotope effect in the superconducting
critical temperature is expected. In the Fe-breathing mode the
linear (in the Fe-displacements) $EPI_{pol}$ coupling vanishes,
while the non-linear (quadratic) one is very strong. The part of
the $EPI_{pol}$ coupling, which is due to the "potential" energy
(the Hubbard U) changes, is responsible for the giant
magneto-elastic effects in $MFe_{2}As_{2}$, $M=Ca$, $Sr$, $Ba$
since it gives much larger contribution to the magnetic pressure
than the band structure effects do. This mechanism is contrary to
the LDA prediction where the magneto-elastic effects are due to
the "kinetic" energy effects, i.e. the changes in the density of
states by the magneto-elastic effects. The proposed $EPI_{pol}$ is
expected to be operative (and strong) in other $Fe$-based
superconductors with electronically polarizable ions such as $Se$,
$Te$, $S$ etc., and in high-temperature superconductors due to the
polarizability of the $O^{2-}$ ions.}

\begin{document}

\maketitle

\section{Introduction}

Recently, superconductivity (SC) with high critical temperature $T_{c}$ was
discovered in several families of $Fe$-pnictides. In the electron (\textit{e}%
) doped ($1111$) system $LaFeAsO_{1-x}F_{x}$ one has $T_{c}\approx 26$ $K$
(and $43$ $K$ at high pressure) \cite{LaFeAsOF}. The record values are $%
T_{c}\approx 55$ $K$ in $SmFeAsO_{1-x}F_{x}$ \cite{SmFeAsO} and $%
T_{c}\approx 56$ K in $Sr_{1-x}Sm_{x}FeAsF$ \cite{Wu}, etc. In the hole ({%
\textit{h}}) doped ($122$) system $Ba_{0.6}K_{0.4}Fe_{2}As_{2}$ one has $%
T_{c}=38$ $K$ \cite{Johrendt}. Other families are reported, such as $MFeAs$
with $M=Li,Na$ and $T_{c}=18$ $K$, and the binary systems $Fe(Te,Se)$ with $%
T_{c}<12$ $K$. A common phase diagram has emerged with: (\textit{i}) the
structural transition appears around $T_{str}=$($140-200$); (\textit{ii})
the SDW-type magnetic ordering occurs at $T_{sdw}\leq T_{str}$, while
superconductivity appears when the SDW ordering vanishes. The latter can be
done either by \textit{e-} or \textit{h}-doping or by applying high pressure.

The important question is - what is the pairing mechanism in $Fe$-pnictides?
The vicinity of these systems to the antiferromagnetic phase was inspiration
for spin fluctuation ($SF$) pairing models with \textit{repulsive
interaction }in the $s$-wave channel. This line of thinking was encouraged
by the small electron-phonon coupling $\lambda _{ep}$ obtained in the LDA\
band structure calculations for $LaO_{1-x}F_{x}FeAs$ \cite{Boeri-Dolgov},
with $\lambda _{ep}^{LDA)}\sim 0.2$ and $T_{c,ep}^{(LDA)}\sim 1$ $K$.
However, this repulsive interaction in the singlet channel might be
effective only if the coupling constant for the scattering of pairs from the
hole to the electron-band - the $hh\leftrightarrows ee$ scattering, is much
larger than the intra-band (repulsive) couplings, i.e. $\left\vert \lambda
_{he}^{sf}\right\vert $ $\gg \left\vert \lambda _{hh}^{sf}\right\vert
,\left\vert \lambda _{ee}^{sf}\right\vert $. In that case the $s_{\pm }$
superconductivity is realized, where the gaps on the \textit{h}- and \textit{%
e}- Fermi surfaces exhibit opposite signs, $sign(\Delta _{h})=-sign(\Delta
_{e})$ \cite{Mazin-interband}. An indirect experimental support for $s_{\pm
} $ pairing would be existence of an very pronounced resonance peak in the
dynamical spin susceptibility \cite{Korshunov} $Im\chi (\omega ,\mathbf{Q})$
at $(\omega /2\Delta _{0})\sim 0.6$ and at the SDW wave vector $\mathbf{Q}$.
The experimental situation is at present unclear since in $%
LaO_{0.87}F_{0.13}FeAs$ \cite{Neutron-res-Lynn} this peak is not observed,
while in \cite{Neutron-Lumsden} it is observed but not very pronounced.
However, several recent NMR measurements of the $T_{1}$-relaxation rate in $%
LaO_{1-x}F_{x}FeAs$, if confirmed, disfavor the $SF$ scenario. First, in
\cite{Buechner} it is reported that $T_{1}$-relaxation rates on the nuclei $%
^{139}$La, $^{57}$Fe, $^{75}$As in $LaO_{1-x}F_{x}FeAs$ scale with one
another, in spite of different form factors in the $q$-space. This means a
lack of any pronounced $q$-structure in $Im\chi (\omega ,\mathbf{q})$, thus
disfavoring the SF pairing mechanism which assume that $Im\chi (\omega ,%
\mathbf{q})$ is strongly peaked at $Q=(1/2,1/2)$. Second, by increasing
doping $x$ in $LaFeAsO_{1-x}F_{x}$ ($0.04\leq x\leq 0.14$) the $T_{1}$%
-relaxation rate of $^{75}$As decreases by almost two orders of magnitude,
while $T_{c}$ is practically unchanged \cite{Nakai-NMR-ASF}, thus again
disfavoring the $SF$ scenario. However, these experiments do not exclude the
first order direct Coulomb interaction as the inter-band pairing mechanism.
In order to increase $T_{c}$ in a multi-band system the \textit{intra-band
pairing} should contribute positively to $T_{c} $. This may happen if the
intra-band pairing potentials are: (\textit{a}) attractive - giving rise to
a conventional s-wave pairing, or (\textit{b}) repulsive and anisotropic -
giving rise to an unconventional pairing. In the latter case the system is
very sensitive to the presence of nonmagnetic impurities. The $s_{\pm }$
pairing is also sensitive to nonmagnetic impurities although in some
occasions they are less detrimental than for unconventional pairing. For
instance, in the unitary limit $s_{\pm }$ is unaffected while in the Born
limit these impurities are pair-breaking for it \cite{Kul-Dolg-imp}, \cite%
{Kulic-1-epl}. The analysis of the resistivity $\rho (T)$ \cite{Kulic-1-epl}
and the upper critical field $H_{c2}(T)$ \cite{Drechsler-Hc2} gives evidence
for appreciable impurity effects in single crystals of the $Fe$-pnictides,
which disfavor any gapless unconventional pairing \cite{Kulic-1-epl}. The
same analysis hints that the \textit{attractive intra-band pairings}, with $%
\lambda _{hh},\lambda _{ee}>0$, are important in the Fe-pnictides which can
be only due to EPI or to an excitonic mechanism \cite{Kulic-1-epl}.

The above results imply a reconsideration of the role of EPI in the
Fe-pnictides. This is supported by the recent neutron scattering experiments
where the structural properties of the $Fe$-pnictides are sensitive to the
magnetic order by showing \textit{giant magneto-elastic effects}. For
instance, in $CaFe_{2}As_{2}$ there is a first-order phase transition from
the orthorhombic to the tetragonal structure under the pressure $P_{c}>0.35$
$GPa$, while at the same time the magnetic order vanishes \cite{Kreyssig}.
Concerning the role of EPI in these materials there were recently two
controversial reports: the first one on a \textit{positive} and \textit{%
large Fe-isotope effect} in $T_{c}$, and surprisingly also in $T_{sdw}$, for
the substitution $^{56}Fe\rightarrow ^{54}Fe$ in $SmFeAsO_{1-x}F_{x}$ and $%
Ba_{1-x}K_{x}Fe_{2}As_{2}$ with $\alpha _{Fe}^{(T_{c})}\approx 0.33-0.4$ and
$\alpha _{Fe}^{(T_{sdw})}\approx 0.35-0.4$ \cite{Liu-isotope}. The second
one \cite{inverse-Iyo} reports, contrary to the first one, on a \textit{%
negative Fe-isotope effect} with $\alpha _{Fe}^{(T_{c})}=-0.18$.
\begin{figure}[tbp]
\resizebox{.4
\textwidth}{!} %{\includegraphics*[width=10cm]{Fe-structure.eps}}
{\includegraphics*[width=8cm]{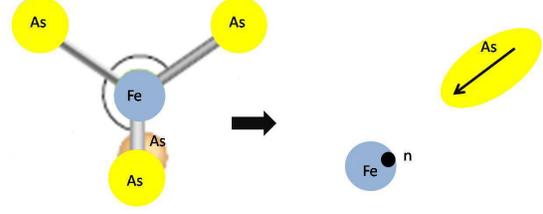}}
\caption{Schematic picture of the polarization of the As
electronic cloud due to charge fluctuations ($n$) on the Fe-ions.
The tiny arrow on the As-ion describes the induced electronic
dipole moment on As.} \label{As-polarization}
\end{figure}

These results point to a rather strong involvement of EPI in magnetic,
elastic and in superconducting properties of Fe-pnictides and to the
importance of many-body effects. In the following we discuss a possibility
for a many-body $EPI$ channel in the $Fe$-pnictides, which is due to the
\textit{large electronic polarizability of the $As$-ions} - we call it $%
EPI_{pol}$. The importance of the electronic polarizability of the ligands
in the screening of the core electrons in the excitonic spectra of halides
and oxides was first recognized and studied by the Sawatzky group in \cite%
{Sawatzky-84} and later on applied on the Mott-Hubbard model for the
screening of the (on-site) atomic Coulomb repulsion by the fast electronic
polarizability processes \cite{Sawatzky-95} - we call it the $SPS$ model
(the Sawatzky polarization screening model). The $SPS$ model was recently
applied to the Fe-pnictides \cite{Sawatzky-08} where the relatively small
Hubbard repulsion on Fe-ions was explained by the screening via the huge
electronic polarizability of the As-ions. Here we argue that if the $SPS$
model is realized in the Fe pnictides then it gives rise inevitably to: (i)
the large $EPI_{pol}$ potential $V_{ep}$ for vibrations of the $As$-ions
(especially in the $As$ $A_{1g}$-modes); (ii) the strong dependence of $%
V_{ep}$ on the $As-Fe$ distance $d_{As-Fe}$ with $V_{ep}\sim d_{As-Fe}^{-4}$%
. As the consequence, the $EPI_{pol}$ coupling $g_{ep}(=\partial
V_{ep}/\partial d_{As-Fe})$ is much larger than the the LDA values since $%
g_{ep}(\sim 20$ $eV/\mathring{A})\gg g_{ep}^{(LDA)}(<1$ $eV/\mathring{A})$,
thus giving an appreciable contribution to the bare $EPI_{pol}$ coupling
constant $\lambda _{ep,A_{1g}}^{0}\approx 1$. The latter may significantly
contribute to the intra-band $s$-wave pairing, thus disfavoring any gapless
unconventional intra-band pairing. In the case of $EPI_{pol}$ an appreciable
positive \textit{As-isotope effect} in $T_{c}$ is expected. Furthermore, $%
EPI_{pol}$ may be responsible for the giant magneto-elastic effects in $%
MFe_{2}As_{2}$, $M=Ca$, $Sr$, $Ba$ . In that sense we construct the
thermodynamic potential $G(S_{Q},\varepsilon ,P)$ as a function of the
strain $\varepsilon $, the SDW magnetic order parameter $S_{Q}$ and the
pressure $P$ by taking into account the effects of $EPI_{pol}$. Based on
this, we argue that the giant magneto-elastic effects in $MFe_{2}As_{2}$, $%
M=Ca$, $Sr$, $Ba$ are dominated by $EPI_{pol}$, since it gives much larger
contribution to the magnetic pressure than that predicted by LDA \cite%
{Yildirim}, \cite{Yndurain}, \cite{Jeschke-Valenti}.

\section{$EPI$ due to the $As$ electronic polarizability}

The electronic part of the Hamiltonian in the SPS model applied to the
Fe-pnictides contains several many-body effects, i.e. $\hat{H}=\hat{H}_{0,e}+%
\hat{H}_{c}+\hat{H}_{H}^{at}+\hat{H}_{core}$. Here, $\hat{H}_{0,e}$ is the
kinetic energy of the conduction electrons, $\hat{H}_{c}$ their long-range
Coulomb interaction, while the Hubbard term $\hat{H}_{H}^{at}$ describes the
\textit{unscreened (on-site) atomic repulsion} $U_{at}$ on the $Fe$-ions. $%
\hat{H}_{core}$ describes the (very) polarizable electronic cores of the $As$%
-ions. Fluctuations of charges on the Fe-ions create an electric field on
the As-ions which induces the polarization $\mathbf{P}_{As}$ of their cores,
thus lowering the energy of the system - see $Fig.1$. By eliminating the
core degrees of freedom the part $H_{H}^{at}+\hat{H}_{cor}$ can be
approximated by the Hamiltonian $\hat{H}^{(pol)}=\hat{H}_{0}^{(pol)}+\hat{H}%
_{H}^{(pol)}+\hat{H}_{c}^{(pol)}$ - see more in \cite{Sawatzky-84}-\cite%
{Sawatzky-08}. This is pure many-body effect and almost not captured by the $%
LDA$ band structure calculations. Applied to the Fe-pnictides it gives%
\begin{equation}
\hat{H}_{0}^{(pol)}=-\sum_{\mathbf{R}_{Fe}}V_{P}(\mathbf{R}_{Fe})\hat{n}_{%
\mathbf{R}_{Fe}}/2,  \label{H0-pol}
\end{equation}%
which renormalizes the local energy on the Fe-ions.

What is important for the magnetic and magneto-elastic properties is tha
this polarization process \textit{renormalizes also the Hubbard on-site
repulsion} giving $\hat{H}_{H}^{(pol)}$

\begin{equation}
\hat{H}_{H}^{(pol)}=\sum_{\mathbf{R}_{Fe}}U_{\mathbf{R}_{Fe}}^{(sc)}\hat{n}_{%
\mathbf{R}_{Fe}\uparrow }\hat{n}_{\mathbf{R}_{Fe}\downarrow },  \label{H-h}
\end{equation}%
with the \textit{screened Hubbard repulsion} $U_{\mathbf{R}%
_{Fe}}^{(sc)}=U_{at}-V_{p}(\mathbf{R}_{Fe})$; $\mathbf{R}_{Fe}$ enumerates
the $Fe$-ions. The screening term $V_{p}(\mathbf{R}_{Fe})$, which is due to
the \textit{electronic polarizability} of the $As$-ions, is given in the
point charge approximation by \cite{Sawatzky-08}
\begin{equation}
V_{p}\approx \sum_{\mathbf{R}_{As}\epsilon n.n.\mathbf{R}_{Fe}}\frac{\alpha
_{As}e^{2}}{\left\vert \mathbf{R}_{Fe}-\mathbf{R}_{As}\right\vert ^{4}},
\label{Vp}
\end{equation}%
where $\mathbf{R}_{As}$ enumerates the $As$-ions which are nearest neighbors
($n.n.$) of a given $Fe$-ion and $\alpha _{As}$ is the electronic
polarizability of the $As$-ion. The physical picture behind the term $%
V_{p}(\equiv 2E_{p})$ is the following: the charge fluctuations on the
neighboring $Fe$-ions cause local electric fields $\mathbf{E}_{loc,As}$ on
the $As$-ions, where $\mathbf{E}_{loc,As}$ polarizes the $As$ electronic
clouds by exciting s-p transitions as shown schematically on Fig.1. The
induced dipole moments lower the energy by an amount $E_{p}\approx \alpha
_{As}\sum_{R_{As}}\mathbf{E}_{loc,R_{As}}^{2}/2$. Due to large $\alpha _{As}$%
, with $\alpha _{As}\sim (10-12)$ $\mathring{A}^{3}$one obtains $V_{p}\sim
10 $ $eV$ \cite{Sawatzky-08}. Since $U_{at}<15e$ $V$ it gives $%
U^{(sc)}\lesssim 3$ $eV$. Since $U^{(sc)}$ is comparable to the band-width $%
W $ this means that the Fe-pnictides are in a moderately correlated regime.

%\begin{figure}[tbp]
%\resizebox{.4
%\textwidth}{!} %{\includegraphics*[width=10cm]{Fe-structure.eps}}
%{\includegraphics*[width=8cm]{As-polarization.eps}}
%\caption{Schematic picture of the polarization of the As
%electronic cloud due to charge fluctuations ($n$) on the Fe-ions.}
%\label{As-polarization}
%\end{figure}

The meaning of $U^{(sc)}=U_{at}-V_{p}$ and $V_{p}$ can be understood by the
following analysisc \cite{Sawatzky-84}-\cite{Sawatzky-08}. The Hubbard
repulsion $U$ in solids is the difference between the ionization energy $I$
and the electron affinity $A$, i.e. $U^{(sc)}=I-A$. In solids the atomic
ionization energy is lowered by $E_{p}$ while the atomic electronic affinity
is increased by $E_{p}$, i.e. $%
U^{(sc)}=I-A=(I_{at}-E_{p})-(A_{at}+E_{p})=U_{at}-V_{p}$. The strong
renormalization of $U_{at}$ to $U^{(sc)}$ is confirmed in the analyzis of
the optical data \cite{Drechsler-optics}, due to the large background
dielectric constant $\varepsilon _{\infty }\approx 12$ which gives $\alpha
_{As}\sim (10-12)$ $\mathring{A}^{3}$ and $U^{(sc)}\approx (1-2)$ $eV$.
(There is also an additional non-local Coulomb term $\hat{H}_{c}^{(pol)}$
due to the $As$ polarization, which depends on $V_{P}(\mathbf{R}_{Fe})$ and
will be not studied here.)

In \cite{Sawatzky-08} a possible excitonic mechanism of superconductivity
was analyzed in the SPS model, while the giant EPI effects were not studied
at all. In the presence of \textit{vibrations} of the $As$- and $Fe$-ions
the change of $\hat{H}^{(pol)}$gives rise to an additional many-body $%
EPI_{pol}$ channel with a possible huge EPI coupling potential $V_{ep}$
where the $EPI_{pol}$ Hamiltonian $\hat{H}_{ep}^{(pol)}(=\hat{H}%
_{0,ep}^{(pol)}+\hat{H}_{H,ep}^{(pol)})$ reads

\begin{equation}
\hat{H}_{ep}^{(pol)}=V_{ep}\sum_{\mathbf{R_{Fe}}}\hat{\varphi}_{\mathbf{R}%
_{Fe}}(\frac{1}{2}\hat{n}_{\mathbf{R}_{Fe}}+\hat{n}_{\mathbf{R}_{Fe}\uparrow
}\hat{n}_{\mathbf{R}_{Fe}\downarrow })  \label{Hep}
\end{equation}%
where $V_{ep}=4V_{p}(\sim 40$ $eV)$ is very large and the dimensionless
phonon operator $\hat{\varphi}_{\mathbf{R}_{Fe}}$ is given by $\hat{\varphi}%
_{\mathbf{R}_{Fe}}\equiv \lbrack V_{p}(\mathbf{R}_{Fe}^{0}+\mathbf{\hat{u}}_{%
\mathbf{R}_{Fe}}-\mathbf{R}_{As}^{0}-\mathbf{\hat{u}}_{\mathbf{R}%
_{As}})-V_{p}(\mathbf{R}_{Fe}^{0}-\mathbf{R}_{As}^{0})]/V_{ep}$ and $\hat{n}%
_{\mathbf{R}_{Fe}}=\hat{n}_{\mathbf{R}_{Fe}\uparrow }+\hat{n}_{\mathbf{R}%
_{Fe}\downarrow }$. Here $\mathbf{\hat{u}}_{\mathbf{R}_{Fe}}$, $\mathbf{\hat{%
u}}_{\mathbf{R}_{As}}$ are the displacement operators of the $Fe$- and $As$%
-ions, respectively. For the \textit{harmonic EPI interaction} one has%
\begin{equation}
\hat{\varphi}_{\mathbf{R}_{Fe}}=\frac{1}{Zd}\sum_{\mathbf{R}_{As}\epsilon
n.n.\mathbf{R}_{Fe}}\mathbf{n}_{As}\cdot (\mathbf{\hat{u}}_{\mathbf{R}_{As}}-%
\mathbf{\hat{u}}_{\mathbf{R}_{Fe}}),  \label{phi}
\end{equation}%
where $\mathbf{n}_{As}=(\mathbf{R}_{Fe}^{0}-\mathbf{R}_{As}^{0})/d_{As-Fe}$.
It is convenient to rewrite the Hubbard Hamiltonian $\hat{H}_{ep}^{(pol)}$in
the rotational invariant form (note $2\hat{n}_{\mathbf{R}_{Fe}\uparrow }\hat{%
n}_{\mathbf{R}_{Fe}\downarrow }=\hat{n}_{\mathbf{R}_{Fe}}-4\mathbf{\hat{S}}_{%
\mathbf{R}_{Fe}}^{2}/3$)
\begin{equation}
\hat{H}_{ep}^{(pol)}=V_{ep}\sum_{\mathbf{R}_{Fe}}\hat{\varphi}_{\mathbf{R}%
_{Fe}}(\hat{n}_{\mathbf{R}_{Fe}}-\lambda \mathbf{\hat{S}}_{\mathbf{R}%
_{Fe}}^{2}),  \label{Hep-rot}
\end{equation}%
where $\lambda =2/3$ and the spin-like operator $\mathbf{\hat{S}}=\hat{c}%
^{\dag }\mathbf{\sigma }_{\alpha \beta }\hat{c}/2$. The $EPI_{pol}$
potential $V_{ep}$ in Eq.(\ref{Hep-rot}) is large due to: (\textit{i}) the
large value of $V_{p}$; (ii) the rapid change of $V_{p}$ with the $Fe$- and $%
As$-distance $d_{Fe-As}(\approx 2.4$ $\mathring{A})$, thus giving very large
induced electron-phonon coupling $g_{ep}^{(pol)}(\equiv
V_{ep}/d_{Fe-As})=(4V_{p})/d_{Fe-As}\sim 16$ $eV/\mathring{A}$. For
comparison, the average $LDA$ coupling constant (coming mainly from the
screened lattice potential) is much smaller, i.e. $g_{ep}^{(LDA)}<1eV/%
\mathring{A}$ \cite{Boeri-Dolgov}. As it is said before, the above model can
be generalized to the multi-orbital case. In that case there is a summation
over the orbital index in Eqs.(\ref{Hep},\ref{Hep-rot}) and there are terms
due to the inter-orbital interaction. For instance, Eq.(\ref{Hep-rot}) reads
$\hat{H}_{ep}^{(pol)}=V_{ep}\sum_{i\mathbf{R}_{Fe}}\hat{\varphi}_{\mathbf{R}%
_{Fe}}\{\hat{n}_{i\mathbf{R}_{Fe}}-\lambda \mathbf{\hat{S}}_{i\mathbf{R}%
_{Fe}}^{2}+\sum_{j\neq i}\hat{n}_{i\mathbf{R}_{Fe}}\hat{n}_{j\mathbf{R}%
_{Fe}}/2\}+\hat{H}_{nloc}$, where $i,j$ enumerates the orbital indices while
all non-local terms mentioned above are stipulated in $\hat{H}_{nloc}$. In
the following we argue that the giant magneto-elastic effects in the
Fe-pnictides are dominated by the $EPI_{pol}$ effects.

\section{Magneto-elastic coupling effects}

\begin{figure}[tbp]
\begin{center}
\resizebox{.3 \textwidth}{!} {\includegraphics*
[width=6cm]{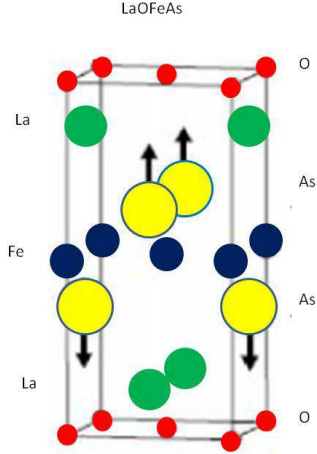}} %{\includegraphics*
%[width=4cm]{Fe-cell.eps}}
\end{center}
\caption{Vibrations of the As-ions in $A_{1g}$ modes of $LaOFeAs$ which produce very strong $%
EPI_{pol}$. Small displacements of the La-ions are not shown.}
\label{A1g mode}
\end{figure}

If the SPS model is realized in the Fe-pnictides then the part of $EPI_{pol}$
- the second term in Eqs. (\ref{Hep},\ref{Hep-rot}), is involved in the
\textit{giant magneto-elastic effects}. It will be shown below that it gives
the largest contribution to them with the following manifestations. First,
in the presence of the SDW magnetic order ($S_{Q}\neq 0$) the Fe-pnictides
show a tendency to increase significantly the $c$-lattice parameter (along
the c-axis) in comparison to the non-magnetic ($S_{Q}=0$) "collapsed"
tetragonal ($cT$) phase, i.e. $c(S_{Q}\neq 0)>c(S_{Q}=0)$. Second, recent
experiments on $MFe_{2}As_{2}$, $M=Ca$, $Sr$, $Ba$ under pressure ($P$) show
the orthorhombic-tetragonal (the $cT$-phase) transition. In $CaFe_{2}As_{2}$
at $P=P_{c}\approx 0.4GPa$ there is a transition from the orthorhombic phase
with the finite SDW order $S_{Q}\neq 0$) to the non-magnetic cT-phase ($%
S_{Q}=0$) in which the superconductivity appears at $T_{c}=12$ $K$ \cite%
{Kreyssig}. This gives evidence for strong and unusual coupling of the
lattice with the magnetic order. It is worth to mention that the
experimental change of volume $(\Delta V/V)\approx 3-4$ $\%$ is explained by
the LDA calculations by assuming much larger critical pressure $P_{c}$ than
in experiments \cite{Jeschke-Valenti}, i.e. $P_{c}^{(LDA)}=5$ $GPa\gg
P_{c}^{(\exp )}\approx 0.4$ $GPa$. The LDA magnetic pressure is given by $%
P_{m}^{(LDA)}\sim -\partial E^{(LDA)}/\partial \varepsilon $ and its small
value causes that $P_{c}^{(LDA)}\gg P_{c}^{(\exp )}$, thus pointing to some
missing many-body effects with much larger \textit{magnetic pressure} ($%
P_{m} $) - see Eq.(\ref{G-strain}) and the analysis below it. In the
following we show, that $EPI_{pol}$ gives very large contribution to $P_{m}$
- we call it the \textit{potential magnetic pressure }$P_{m}^{(p)}$, thus
lowering the value of $P_{c} $.

The magneto-elastic properties of the tetragonal-orthorhombic Fe-pnictides
are complicated since they depend on six elastic constants, on several other
magneto-elastic constants and on the magnetic free-energy. However, for our
purpose to show that $P_{m}^{(p)}\gg $ $P_{m}^{(LDA)}$ a simplified
phenomenological mean-field approach is satisfactory. For the further
analysis the important experimental fact is the \textit{decrease of the} $Fe-
$ $As$ \textit{distance} $d_{Fe-As}$ by $1$ $\%$ in the first-order phase
transition, i.e. $\epsilon _{d}=(\delta d_{Fe-As}/d_{Fe-As})=0.01$. The
latter lowers the "potential" energy $U_{\mathbf{R}_{Fe}}^{(sc)}$
significantly which leads to the giant magneto-elastic effects. To estimate
the effect we use the experimental fact that at the critical pressure $P_{c}$
the volume is changed by $3-4$ $\%$ \cite{Kreyssig}, while $\epsilon _{d}$
is only a part of the total strain $\varepsilon (\equiv \delta
V/V)=\varepsilon _{xx}+\varepsilon _{yy}+\varepsilon _{zz}$, i.e. $%
\varepsilon _{d}\approx \varepsilon /r$ with $r\sim 3-4$. According to Eqs.(%
\ref{Hep}-\ref{Hep-rot}) one has $U^{(sc)}(\delta d)=U^{(sc)}+V_{ep}\epsilon
_{d}\approx U+4V_{p}\epsilon /r$ (note $V_{ep}=4V_{p}$, $V_{p}\sim 10$ $eV$%
), i.e. the "potential" energy is significantly decreased under the pressure
(for $\epsilon <0$). The weak SDW magnetic order (with $S_{Q}\neq 0$) can be
qualitatively analyzed in the Hartree-Fock (or Stoner-like) approximation
with the bare spin susceptibility $\chi _{0}(Q)=N(0)f(Q)$, where $N(0)$ is
the density of states at the Fermi level and $f(Q)$ describes the momentum
dependence of the bare spin susceptibility. For $U^{(sc)}(P<P_{c})\chi
_{0}(Q)>1$ it gives finite value for $S_{Q}$. The phenomenological Gibbs
energy $G(S_{Q},\varepsilon ,P)$ includes: (\textit{i}) the complicated
elastic properties of the lattice approximately via the effective
compressibility, (\textit{ii}) the SDW magnetism, (\textit{iii}) the
magnetostriction and (\textit{iv}) the effects of the applied pressure $P$
\begin{equation}
G(S_{Q},\varepsilon ,P)\approx \frac{\varepsilon ^{2}}{2\kappa _{eff}}%
+P\varepsilon -\frac{a(\varepsilon )}{2}S_{Q}^{2}+\frac{b}{4}S_{Q}^{4}+\frac{%
c}{6}S_{Q}^{6},  \label{G-strain}
\end{equation}%
where $a(\varepsilon )=U^{(sc)}(\varepsilon )-\chi _{0}^{-1}(Q,\varepsilon
,T)$. The first term in $G$ is an effective lattice elastic energy with the
effective compressibility $\kappa _{eff}$, while the second one is due to
the work by the pressure $P$. The magnetic part of $G$ contains higher order
terms in $S_{Q}^{2}$ which are proportional to $b$ and $c$ since the
transition may be of the first order if $b<0$, or if $b>0$ for $%
b_{r}=b-\kappa _{eff}(\gamma _{k}+\gamma _{p})^{2}/2<0$ (see below), where $%
b_{r}$ is the renormalized fourth-order coefficient in $G$. Note, that here
we consider only effects of the \textit{homogenous strain} on the magnetic
transition in weak ferromagnets ($S_{Q}^{2}\ll 1$) in the mean-field
approximation, which can take place even very far from the critical
temperature $T_{sdw}$. In the following we omit the temperature effects due
to the strain fluctuations which are pronounced very near T$_{sdw}$ - the
Larkin-Pikin mechanism \cite{Larkin-Pikin}. The latter may also lead to the
first order transition due to the interaction of the magnetic order \cite%
{Larkin-Pikin} with the share-strain fluctuations which is studied in \cite%
{Barzykin}. For small strain $\varepsilon \ll 1$ one has
\begin{equation}
a(\varepsilon )\approx a_{0}+(\gamma _{k}+\gamma _{p})\varepsilon ,
\label{a-eps}
\end{equation}%
where $a_{0}=[U^{(sc)}-\chi _{0}^{-1}(Q,0,T)]$. The "kinetic" energy
contribution to the magnetic pressure is proportional to $\gamma _{k}=\chi
_{0}^{-1}(Q,0)d\ln \chi _{0}(Q,\varepsilon )/d\varepsilon $, while the
"potential" energy contribution to $\gamma _{p}=V_{ep}/r$. After the
minimization of $G(S_{Q},\varepsilon ,P)$ (with respect to $S_{Q}$ and $%
\varepsilon $) and for $b_{r}<0$ the \textit{first order phase transition}
is realized at the critical pressure $P_{c}$

\begin{equation}
P_{c}=\frac{1}{(\gamma _{k}+\gamma _{p})\kappa _{eff}}\left( a_{0}+\frac{%
3b_{ren}^{2}}{16c}\right) .  \label{Pc}
\end{equation}%
The SDW order parameter jumps from $S_{Q}\neq (3\left\vert b_{r}\right\vert
/4c)^{1/2}$ at $P=P_{c}-0$ to $S_{Q}=0$ at $P=P_{c}+0$. In the case of the
weak first order phase transition one has $b_{ren}^{2}\ll 4c$ which gives $%
P_{c}\approx a_{0}/(\gamma _{k}+\gamma _{p})\kappa _{eff}$, as well as for
the \textit{second order phase transition} with $b_{r}>0$ (where $S_{Q}=0$
for $P=P_{c}-0$). Note, the the "kinetic" energy term $\gamma _{k}$ in Eq. (%
\ref{Pc}) is contained in the $LDA$ band structure calculations \cite%
{Yildirim}, \cite{Yndurain}, \cite{Jeschke-Valenti}, where it is solely
responsible for the magneto-elastic effects. However, the "potential" term $%
\gamma _{p}$ which is due to the many-body $EPI_{pol}$ effects is absent in
the LDA calculations. This is the main reason that LDA gives too large value
for $P_{c}$, i.e. $P_{c}^{(LDA)}\sim 15P_{c}^{(\exp )}$ \cite%
{Jeschke-Valenti}, which implies that $\gamma _{p}\gg \gamma _{k}(\approx
\gamma _{LDA})$ is realized. Indeed, in most itinerant magnets one has $%
\gamma _{k}<2N^{-1}(0)$ and one expects similar value in the $Fe$-pnictides.
Since $V_{ep}(\sim 40$ $eV)$ is very large one has $(\gamma _{p}/\gamma
_{k})\sim 10$. The latter result implies much smaller critical pressure $%
P_{c}\ll P_{c}^{(LDA)}$ than the LDA predicts. This means that the giant
magneto-elastic effects in the Fe-pnictides are dominated by the many-body
"potential" energy effects which are contained in $EPI_{pol}$. The
experimental situation concerning the type of the phase transition (first-
or second-order) in $MFe_{2}As_{2}$, $M=Ca$, $Sr$, $Ba$ is still not
resolved. Namely, for $Ca$, $Sr$ the case $b_{r}<0$ is realized while in $Ba$
both cases $b_{r}>0$ and $b_{r}<0$ might be realized \cite{Kreyssig}. In
spite of our limited phenomenological approach some semi-quantitative
estimations are possible. For instance, one expects $\kappa _{eff}<2\times
10^{-2}$ $(GPa)^{-1}$ \cite{Shein} which gives $P_{c}\kappa _{eff}<10^{-2}$
and $a_{0}/(\gamma _{k}+\gamma _{p})<10^{-2}$, $a_{0}/\gamma _{p}<10^{-2}$.
The latter condition is a constraint for the microscopic models for the SDW
order parameter and the $EPI_{pol}$ coupling.

\section{Possible contribution of $EPI_{epl}$ to pairing}

At present there is a confusion regarding the role of EPI in the
superconducting pairing of the Fe-pnictides. In spite of numerous
experiments which give appreciable evidence against gapless unconventional
pairing and that the antiferromagnetic spin fluctuations are moderate in
superconducting samples, the proponents of the spin fluctuation mechanism
rely their believe on the small $EPI$ calculated in the $LDA$ calculations
with $\lambda _{ep}\approx 0.2$ and $T_{c}\sim 1$ $K$ \cite{Boeri-Dolgov}.
However, like in the case of high-temperature superconductors (HTSC), heavy
fermions and organic superconductors, etc. the LDA calculations fail
whenever many body effects play important role. For instance, in case of
HTSC numerous tunnelling, transport and neutron scattering measurements
point to a strong EPI, while various LDA calculations furnished diversity of
values for $\lambda _{ep}$, from $0.2-1$ - see discussion in \cite%
{Kulic-review}. The situation is similar in Fe-pnictides where LDA cannot
explain quantitatively even the magnetic and structural properties, let
alone the EPI effects and superconductivity. Here, we give some qualitative
estimate of the \textit{polarization induced} EPI which is described by $%
\hat{H}_{ep}^{(pol)}$ in Eq.(\ref{Hep-rot}) and not contained in LDA.

The strongest $EPI_{pol}$ coupling is for the $A_{1g}$-modes (there are at
least two such modes) where the $As$-ions vibrate along the $c$-axis as
shown in Fig.2. These modes contribute mainly to the \textit{intra-band
pairing}. At present we can only estimate the intra-band ($i=e,h$) bare
coupling constants $\lambda
_{ep,A_{1g}}^{0,i}=2N_{i}(0)g_{A_{1g}}^{2}/\omega _{A_{1g}}$ with $%
g_{A_{1g}}^{2}\approx V_{ep}^{2}\cos ^{2}\theta \left\langle \hat{u}%
_{As}^{2}\right\rangle /d_{Fe-As}^{2}$. Here, $\omega _{A_{1g}}$ is the
energy of an $A_{1g}$-mode, $N_{i}(0)$ is the (intra-band) density of
states, $\left\langle \hat{u}_{A_{1g}}^{2}\right\rangle $ is the average of
the quadratic As-displacement, $\theta $ is the angle between $\mathbf{n}%
_{A} $ and the $c$-axis. In the approximation of one-phonon processes one
has $\left\langle \hat{u}_{A_{1g}}^{2}\right\rangle \approx \hbar
^{2}/2M_{As}\omega _{A_{1g}}$ where $M_{As}$ is the atomic mass of the
As-ion. For $\omega _{A_{1g}}\approx 25$ $meV$ \cite{Hadjiev} and $%
N_{i}(0)\approx 0.5$ $states/eV\cdot spin$ (note that $N_{e}(0)\sim
N_{h}(0)\sim N(0)$) one obtains $\lambda _{ep,A_{1g}}^{0,i}\approx 0.7$. The
estimated $\lambda _{ep,A_{1g}}^{0,i}$is rather large giving a hope that the
real (renormalized) coupling $\lambda _{ep,A_{1g}}(<\lambda
_{ep,A_{1g}}^{0,i})$, might be also appreciable. In order to calculate $%
\lambda _{ep,A_{1g}}^{i}$ one should correctly take into account the matrix
elements of $g_{A_{1g}}$ in the band-basis, screening effects, etc. Even if $%
\lambda _{ep,A_{1g}}^{i}\approx 0.3$ this would give an appreciable total
intra-band coupling constant $\lambda _{ep,A_{1g}}=2\sum_{i}\lambda
_{ep,A_{1g}}^{i}\sim 1$ thus contributing significantly to the effective
pairing constant $\lambda _{eff}$ - see \cite{Kulic-1-epl}.

In \cite{Eschrig} was proposed that the Fe-breathing mode may give
appreciable EPI due to the change of the Fe-Fe hopping energy in this mode.
However, the latter effects are not considered here. In that respect, we
point out that the linear (in the displacement $\mathbf{u}_{Fe}$)
contribution of the Fe-breathing mode to $EPI_{pol}$ vanishes due to the
symmetry reasons - see Fig.2, but this mode gives very \textit{large
nonlinear }$EPI_{pol}$\textit{\ coupling} (with $\hat{\varphi}_{\mathbf{R}%
_{Fe}}\sim $ $\mathbf{u}_{Fe}^{2}$) due to the large value of $\partial
^{2}V_{ep}/\partial d_{Fe-As}^{2}$ - see Eq.(\ref{Vp}). The large nonlinear $%
EPI_{pol}$ coupling is the strongest in the Fe-breathing mode and its effect
on superconductivity and magneto-elastic properties might be also important.

\section{Conclusions}

We have analyzed a possibility for an appreciable many-body contribution to
the electron-phonon interaction ($EPI$) in the Fe-pnictides, which is due to
the electronic polarization of the $As$- ions in the presence of charge
fluctuations on the Fe-ions which is proposed in \cite{Sawatzky-08}. In that
case the polarization induced $EPI_{pol}$ potential $V_{ep}\sim 40$ $eV$ is
especially large for vibrations of the $As$-ions and depends strongly on the
As-Fe distance $d_{As-Fe}$, i.e. $V_{ep}\sim d_{As-Fe}^{-4}$. The
corresponding $EPI_{pol}$ coupling $g_{ep}^{pol}(=\partial V_{ep}/\partial
d_{As-Fe})$ is much larger than the one obtained in the $LDA$ band structure
calculations, i.e. $g_{ep}^{pol}\gg g_{ep}^{(LDA)}$ with $g_{ep}^{pol}(\sim
16$ $eV/\mathring{A})\gg g_{ep}^{(LDA)}(<1$ $eV/\mathring{A})$. The latter
may significantly contribute to the \textit{intra-band s-wave pairing} with
the bare coupling constant $\lambda _{ep,A_{1g}}^{0}\sim 1$, thus
disfavoring any gapless unconventional intra-band superconductivity. As the
result an appreciable positive As-isotope effect in the superconducting
critical temperature is expected. The $EPI_{pol}$ channel gives very strong
nonlinear coupling in the Fe-breathing mode, which might be important for
superconductivity and additional magneto-elastic effect.

The electron-lattice coupling $EPI_{pol}$ gives much larger contribution to
the magnetic pressure than the LDA calculations predict, thus giving rise to
the giant magneto-elastic effects in the transition from the magnetic to the
non-magnetic state under pressure in $MFe_{2}As_{2}$, $M=Ca$, $Sr$, $Ba$.
The critical pressure $P_{c}$ for the magnetic and structural phase
transition is dominated by the "potential" energy contribution $EPI_{pol}$
in the magnetic pressure, while the $LDA$ "kinetic" energy effects are much
smaller. $EPI_{pol}$ is expected to be important in other $Fe$-based
superconductors which contain highly electronically polarizable ions such as
$Se$, $Te$, $S$ etc. It might be operative also for some oxygen modes in
high-temperature superconductors due to the appreciable electronic
polarizability of the $O^{2-}$-ions.

\end{document}